\documentclass{mn2e}
\usepackage{epsf,times}
\newcommand{\etal}{{et al}\/.}

\begin{document}
\title[IC emission from cluster radio galaxies]{Searching for the inverse-Compton
  emission from bright cluster-centre radio galaxies}
\author[M.J.~Hardcastle \& J.H.~Croston]{M.J.\ Hardcastle$^1$\thanks{E-mail: m.j.hardcastle@herts.ac.uk} and
  J.H.\ Croston$^{2,1}$\\$^1$ School of Physics,
  Astronomy and Mathematics, University of
Hertfordshire, College Lane, Hatfield, Hertfordshire AL10 9AB\\$^2$
School of Physics and Astronomy, University of Southampton,
Southampton, SO17 1BJ\\}
\maketitle
\begin{abstract}
We use deep archival {\it Chandra} and {\it XMM-Newton} observations
of three of the brightest cluster-centre radio galaxies in the sky,
Cygnus A, Hercules A and Hydra A, to search for inverse-Compton
emission from the population of electrons responsible for the
low-frequency radio emission. Using simulated observations, we derive
robust estimates for the uncertainties on the normalization of an
inverse-Compton component in the presence of the variations in
background thermal temperature actually seen in our target objects.
Using these, together with the pressures external to the lobes, we are
able to place interesting upper limits on the fraction of the energy
density in the lobes of Hydra A and Her A that can be provided by a
population of relativistic electrons with standard properties,
assuming that the magnetic field is not dominant; these
limits are consistent with the long-standing idea that the energy
density in these lobes is dominated by a non-radiating particle
population. In Cygnus A, we find evidence in the spectra for an
additional hard component over and above the expected thermal
emission, which is plausibly a detection of inverse-Compton emission;
even in this case, though, some additional non-radiating particles and/or
a departure from our standard assumptions on the electron spectrum are
necessary to allow pressure balance at the mid-point of the lobes.
As this is not the case in other Fanaroff-Riley class II radio
galaxies, we suggest that the rich environment of Cygnus A may have
some effect on its lobe particle content.
\end{abstract}

\section{Introduction}
\label{intro}

Inverse-Compton emission appears in many powerful, Fanaroff \& Riley
(1974) class II (hereafter FRII) radio galaxies (e.g. Hardcastle
\etal\ 2002; Isobe \etal\ 2002; Kataoka \& Stawarz 2005; Croston
\etal\ 2005; Hardcastle \& Croston 2005; Isobe \etal\ 2005; Konar
\etal\ 2009; Isobe \etal\ 2009) to be the dominant X-ray emission
mechanism on the scales of the radio lobes, readily detectable in
X-ray observations with moderate sensitivity using {\it Chandra}, {\it
  XMM-Newton}, or, for giant sources, {\it Suzaku}. The close
similarity between the radio and X-ray structures, the agreement
between predicted and observed flux levels, and, where it can be
measured, the spectrum of the emission are all consistent with the
inverse-Compton model (e.g. Hardcastle \& Croston 2005). This allows
us to use lobe-related X-ray emission as a very sensitive probe of the
energy density in low-energy electrons within the lobes, and, for
example, to search for variations in the magnetic field and electron
energy spectrum as a function of position (e.g.\ Isobe \etal\ 2002;
Hardcastle \& Croston 2005; Goodger \etal\ 2008). Comparison of the
energy densities in electrons and magnetic field in the lobes with the
external pressures (e.g.\ Hardcastle \etal\ 2002; Croston \etal\ 2004;
Konar \etal\ 2009) suggests that the lobes of these inverse-Compton
detected sources do not contain an energetically dominant population
of non-radiating particles such as protons.

Less attention has been paid so far to the implications of
inverse-Compton emission in the large number of objects in which it
has so far {\it not} been detected. This includes, at the time of
writing, all classical twin-jet FRI sources, as well as FRII radio
galaxies in environments much richer than a poor cluster, including
some well-studied objects such as Cygnus A (Wilson \etal\ 2000),
3C\,123 (Hardcastle \etal\ 2001), 3C\,295 (Harris \etal\ 2000) and
3C\,220.1 (Worrall \etal\ 2001). In such objects it is assumed that
the inverse-Compton emission is present at some level (as it must be,
since inverse-Compton emission is a required counterpart of the
observed synchrotron emission) but is not readily visible, even in
sources in which it might otherwise be predicted to be seen, because
of the bright thermal bremsstrahlung from the group or cluster
environment.

\begin{table*}
\caption{Sources studied in this paper. $S_{\rm 178}$ is the 178-MHz
  flux density on the Baars \etal\ (1977) scale; for Hydra A this is
  interpolated from the measurements at 160 and 750 MHz quoted by Kuhr
  \etal\ (1981), for Her A it is taken from Spinrad \etal\ (1985) and
  for Cyg A it is the Baars \etal\ value. $\alpha$ is the
  low-frequency spectral index, determined from the Kuhr \etal\ flux
  densities for Hydra A and taken from Spinrad \etal\ (1985)
  otherwise. LAS denotes the largest angular size of the radio
  structures (measured at low frequencies) and LLS is the
  corresponding linear size. The column densities quoted are the
  Galactic values from the compilations of Dickey \& Lockman (1990: NRAO)
  and Kalberla \etal\ (2005: LAB); see the text for discussion.}
\label{sources}
\begin{tabular}{llrrrrrrrr}
\hline
Source&3C name&$z$&$S_{178}$&$\alpha$&$L_{178}$&LAS&LLS&\multicolumn{2}{c}{$N_{\rm H}$ (cm$^{-2}$)}\\
&&& (Jy)&&(W Hz$^{-1}$
sr$^{-1}$)&(arcsec)&(kpc)&(NRAO)&(LAB)\\
\hline
Hydra A&3C\,218&0.054&228&0.70&$1.2 \times 10^{26}$&470&493&$4.8\times
10^{20}$&$4.24 \times 10^{20}$\\
Hercules A&3C\,348&0.154&351&1.00&$1.8 \times 10^{27}$&193&515&$6.3
\times 10^{20}$&$5.95 \times 10^{20}$\\
Cygnus A&3C\,405&0.0565&9660&0.74&$5.8 \times 10^{27}$&129&141&$30.6
\times 10^{20}$&$23.6 \times 10^{20}$\\
\hline
\end{tabular}
\end{table*}

Superficially, this explanation makes some physical sense
in the case of the FRIIs in rich cluster environments. If we assume
[motivated by the results of Hardcastle \etal\ (2002) and Croston
\etal\ (2005)] that the internal pressure in these objects is
dominated by electrons and magnetic field, with the electrons being
energetically dominant by some factor $D$, i.e.
\[
\int E N(E) {\rm d}E = D {{B^2}\over{2\mu_0}}
\]
(where $E$ is the electron energy, $N(E)$ is the electron energy
spectrum, $B$ is the magnetic field strength and $\mu_0$ is the
magnetic constant) and moreover that the electron energy spectra at the (numerically
dominant) low energies are similar in all sources, then for pressure
balance with an external medium with pressure $p_{\rm ext}$ we have
\[
{1\over 3}(1 + 1/D) \int E N(E) {\rm d}E \approx p_{\rm ext}
\]
and thus the normalization of the electron energy spectrum scales
linearly with $p_{\rm ext}$. The volume emissivity for inverse-Compton
scattering of a given photon field goes as the number density of
electrons of the appropriate energy (with Lorentz factors $\gamma
\approx 1000$ for scattering of CMB photons into the X-ray at $z=0$)
and so the observed inverse-Compton surface brightness goes as the
line-of-sight depth through the lobe, $L$, times $p_{\rm ext}$. On the
other hand for thermal bremsstrahlung we have $p_{\rm ext} \approx
(n_p + n_e)kT$, where $T$ is the temperature and $n_p$ and $n_e$ are
the proton and electron densities respectively, we know that $n_e =
\eta n_p$ where $\eta$ is a constant of order unity, and we have a
volume emissivity that goes as $n_pn_eT^{1/2}$. It follows that the
volume emissivity scales as as $L'p_{\rm ext}^{2}/T^{3/2}$, where $L'$
is a characteristic line-of-sight depth that depends on the spatial
structure of the hot gas but that will be of the same order of
magnitude as $L$. Neglecting temperature effects (i.e. making the
observationally motivated assumption that $T$ varies comparatively
little over the typical environments of radio galaxies, and that
differences in pressure are dominated by differences in density) we
see that there is a stronger dependence of the bremsstrahlung
emissivity on $p_{\rm ext}$ than there is for the inverse-Compton
emissivity; thus, at any given redshift (which fixes both the photon
number density of the CMB and the rest-frame waveband in which we are
operating) we expect that there will be some external pressure $p_{\rm
  ext}$ at which thermal bremsstrahlung will come to dominate the
observed X-ray surface brightness, and therefore will prevent a
straightforward detection of inverse-Compton emission.

It is much less obvious that this type of argument can explain the
lack of observed inverse-Compton emission in the lobes of FRI sources,
which are often found in group environments very similar to those of
low-power FRIIs and so might be expected to have similar levels of
inverse-Compton detectability. However, in these sources we already
know that the model discussed above is not valid, in the sense that
the internal pressure of the lobes is almost certainly not supplied by
electrons and magnetic fields close to equipartition (e.g. Croston
\etal\ 2008 and references therein). In this case {\it quantitative}
limits on inverse-Compton emission from the lobes would put some
constraints on the contribution of electrons to the total internal
pressure. In the past we have used the lack of detected
inverse-Compton emission to rule out models in which the energetics of
FRI lobes are {\it dominated} by electrons (e.g. Hardcastle, Worrall
\& Birkinshaw 1998b, Croston \etal\ 2003) but little work has been
done beyond that except in some rather specialized cases (e.g. Jetha
\etal\ 2008).

In this paper we explore the constraints that can be put on the level
of lobe inverse-Compton emission that is present in systems, both FRI
and FRII, in which it cannot readily be detected by eye. We show that
it is possible to place interesting limits on the inverse-Compton
emission from powerful FRIs and intermediate-morphology sources even
in rich environments, supporting a model in which electrons provide a
negligible fraction of the internal pressure in the large-scale lobes
of these objects. We argue that there is a significant detection of
inverse-Compton emission from the lobes of the powerful FRII Cygnus A,
although the simulation-derived uncertainties on its normalization are
large. Our results may shed some light on the mechanism by which the
large-scale components of radio sources in rich environments come to
be dominated by a non-radiating particle population.

In what follows we use a concordance cosmology with $H_0 = 70$
km s$^{-1}$ Mpc$^{-1}$, $\Omega_{\rm m} = 0.3$ and $\Omega_\Lambda =
0.7$. Spectral indices $\alpha$ are the energy
indices and are defined in the sense that flux $\propto
\nu^{-\alpha}$; the photon index $\Gamma = 1 + \alpha$.

\section{Data}

Our aim is to select sources that sample a range of radio
morphological or luminosity classes, from twin-jet to classical FRII
sources, and that a priori plausibly should have a strong
inverse-Compton signal. In the past (Hardcastle \& Croston 2005,
Goodger \etal\ 2008) we have based our selection on low-frequency
radio flux, on the basis that this is a good indicator of the presence
of the electrons required to scatter microwave-background photons into
the X-ray band. If we look at the brightest ten or so extragalactic
objects in the sky at 100-MHz frequencies, the radio galaxies that do
not yet have lobe inverse-Compton detections (setting aside the
special case of Cen A; see Hardcastle \etal\ 2009 for a discussion of
the existing and future limits on inverse-Compton emission from this
object) are Cygnus A (3C\,405), Virgo A (M87, 3C\,274), Hydra A
(3C\,218) and Hercules A (3C\,348). Of these, we exclude M87 because
of the very complex relationship between its thermal X-ray emission
and large-scale low-frequency radio structure (e.g. Simionescu
\etal\ 2007) leaving us with Cyg A, Her A and Hydra A (Table
\ref{sources}). Conveniently these three are, respectively, an
archetypal FRII source, an object with intermediate FRI/FRII
morphology (e.g. Gizani \& Leahy 2003) and a plumed
FRI, probably a wide-angle tail (by the definition of Leahy 1993),
although it is worth noting (see Table \ref{sources}) that all three
lie well above the original FRI/FRII luminosity break ($L_{\rm 178} =
5 \times 10^{24}$ W Hz$^{-1}$ sr$^{-1}$). All three of these radio
galaxies lie in rich cluster environments and have X-ray emission that
is clearly dominated by thermal bremsstrahlung and that does not
display a particular complex relationship to the radio source. Thus
they provide both a scientifically interesting sample and an
appropriate starting point for a study of inverse-Compton emission in
the presence of dominant thermal emission.

\begin{table*}
\caption{X-ray data used in the analysis. Livetime quoted is after
  filtering for any intervals of high background; for the {\it XMM}
  data the three times quoted are for the pn, MOS1 and MOS2 cameras respectively.}
\label{observations}
\begin{tabular}{lllllr}
\hline
Source&Observatory&Date&Observation ID&Detector&Livetime (s)\\
\hline
Hydra A&{\it XMM-Newton}&2007 May 11&0504260101&EPIC&56801, 91356, 91092\\
Hercules A&{\it Chandra}&2005 May 09&5796&ACIS-S&47544\\
&&2005 May 25&6257&ACIS-S&49518\\
Cygnus A&{\it Chandra}&2000 May 21&360&ACIS-S&34720\\
&&2005 Feb 22&5830&ACIS-I&23454\\
&&2005 Feb 16&5831&ACIS-I&51093\\
&&2005 Feb 15&6225&ACIS-I&24308\\
&&2005 Feb 19&6226&ACIS-I&23833\\
&&2005 Feb 25&6228&ACIS-I&16038\\
&&2005 Feb 23&6229&ACIS-I&22753\\
&&2005 Feb 21&6250&ACIS-I&6957\\
&&2005 Sep 07&6252&ACIS-I&29653\\
\hline
\end{tabular}
\end{table*}

Large amounts of archival X-ray data are available for all three of
these sources (Table \ref{observations}). For Hydra A, with a largest
angular size around 7 arcmin, we have chosen to use the {\it
  XMM-Newton} dataset described by Simionescu \etal\ (2009), although
there is also deep {\it Chandra} data (Wise \etal\ 2007). For Her A we
use unpublished archival {\it Chandra} data which go somewhat deeper
than the observations described by Nulsen \etal\ (2005). Finally, for
Cygnus A, we use the extensive {\it Chandra} dataset described by e.g.
Wilson, Smith \& Young (2006).

The X-ray data were reprocessed from the archive in the standard
manner using {\sc ciao} 3.4 for the {\it Chandra} observations and
{\sc sas} 8.0 for {\it XMM}. For the {\it XMM} data we filtered using
the standard flags. As the observation was somewhat affected by
flaring, we filtered data in the standard manner, using a light curve
for the whole FOV of each camera in the nominal energy range 10--15 keV and
excluding intervals where the count rate in this band exceeded 1 count
s$^{-1}$ (pn) or 0.35 s$^{-1}$ (MOS). No time filtering was necessary
for the {\it Chandra} data. Spectra were extracted using {\it
  especget} ({\it XMM}) or {\it specextract} ({\it Chandra}) and
grouped to have either 40 ({\it XMM}) or 20 ({\it Chandra}) counts per
bin after background subtraction (the larger binning factor used for the
{\it XMM} data was intended to compensate for the lower signal-to-noise
of the regions of interest). Throughout the paper spectral fitting was
carried out using {\sc xspec} 12.

\begin{figure*}
\epsfxsize 10cm
\epsfbox{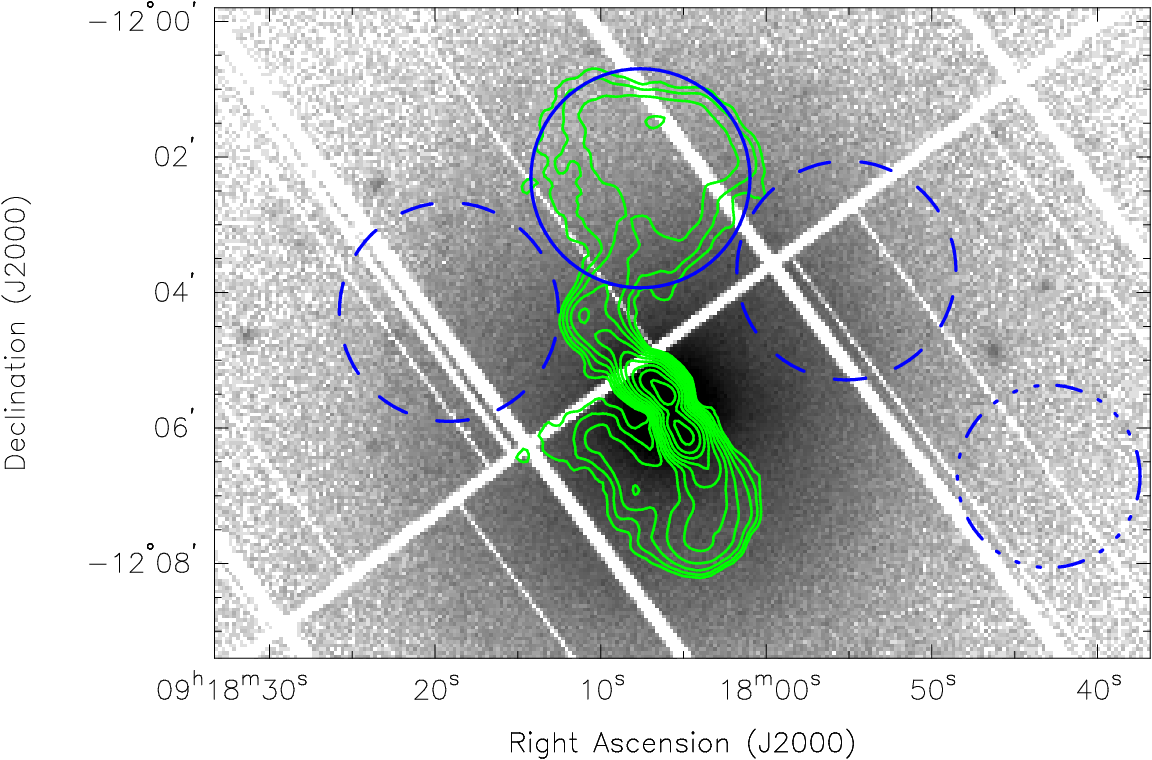}
\epsfxsize 10cm
\epsfbox{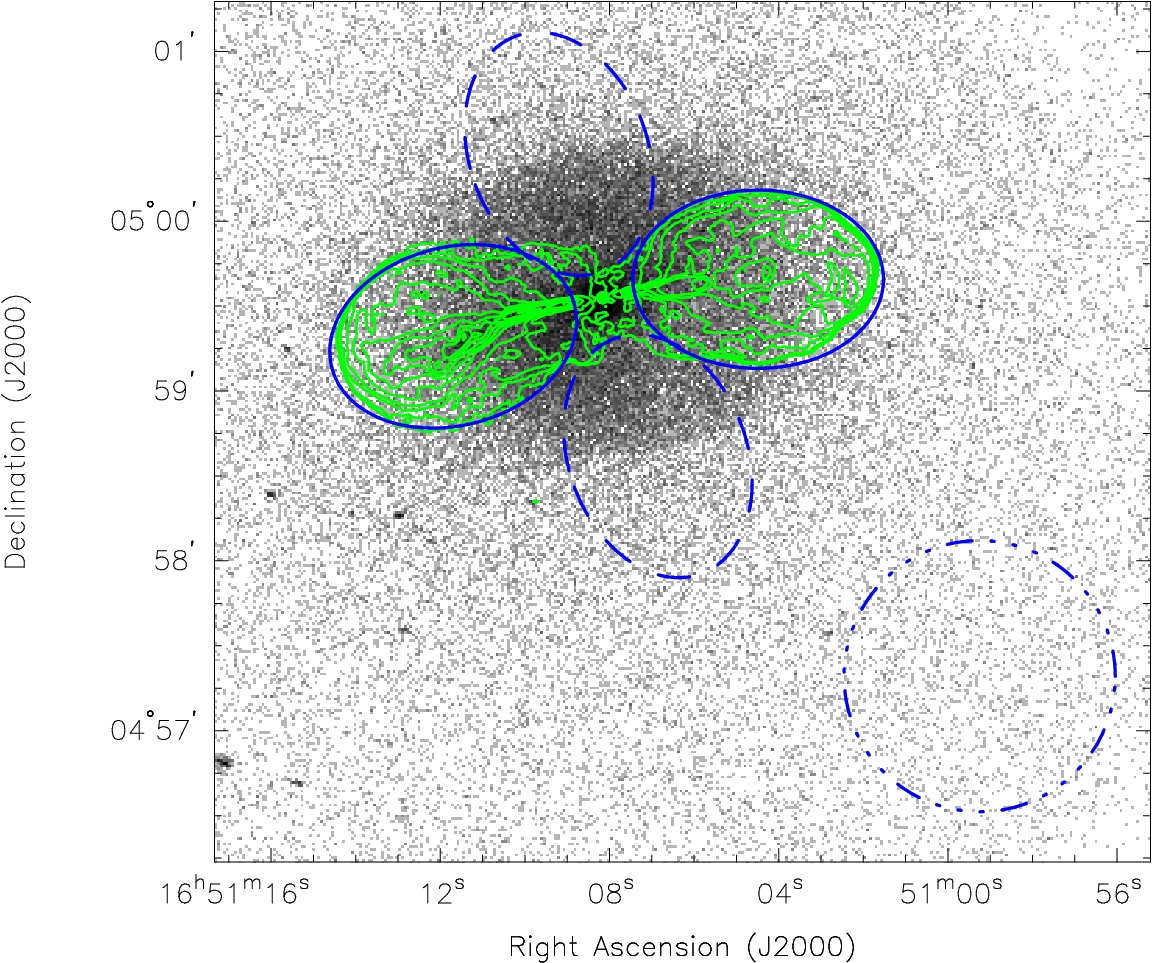}
\epsfxsize 10cm
\epsfbox{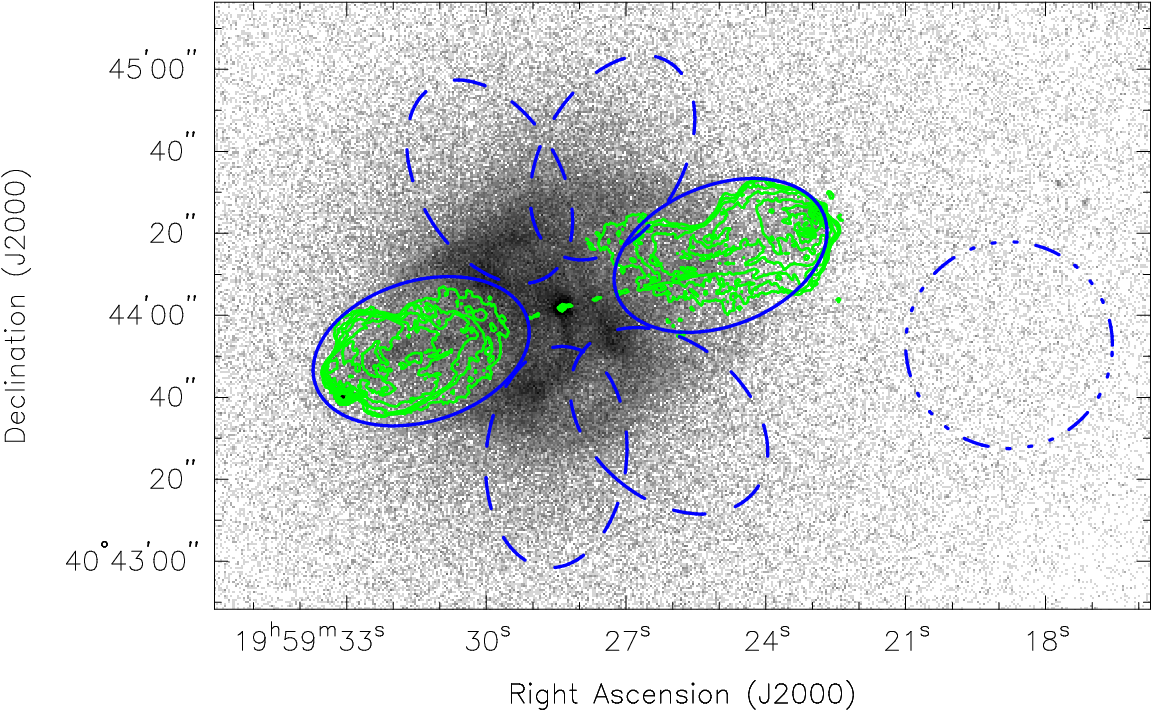}
\caption{The target objects. Top: Hydra A. Middle: Her A. Bottom:
  Cygnus A. Logarithmic grayscales show X-ray counts in the 0.5-5.0
  keV band from the {\it XMM} pn
  observation (Hydra A) and from the merged {\it Chandra} observations
(Her A and Cyg A). Radio contours are logarithmic, increasing by a
  factor 2, and show respectively the 330-MHz, 1.4 GHz and 5-GHz
  images described in the text. The lobe, off-lobe and background
  regions are shown as blue circles or ellipses; solid lines indicate
  lobe regions, dashed off-lobe regions and dot-dashed background.}
\label{picture}
\end{figure*}

Radio data are necessary to interpret inverse-Compton detections or
upper limits. For Cygnus A, we used the 5-GHz radio map of Carilli
\etal\ (1991), obtained from the NCSA Astronomy Digital Image
Library\footnote{http://imagelib.ncsa.uiuc.edu/imagelib.html}. For
Hydra A, we used the 74-MHz and 330-MHz images of Lane \etal\ (2004),
kindly supplied by Wendy Lane, while for Her A the maps used are the
1.4 and 5-GHz maps of Gizani \& Leahy (2003), kindly provided by Paddy
Leahy. X-ray images of the three objects investigated in this paper, with
radio contours and lobe regions overlaid, are shown in Fig.\ \ref{picture}.

Finally, for spectral fitting, especially of complex models, it is
necessary to have an estimate of the level of Galactic absorption.
When searching for weak non-thermal inverse-Compton emission in the
presence of strong thermal emission, an incorrect estimate of the
absorbing column can lead to false positives (or negatives), while
leaving it free in the fits is also likely to have bad effects. There
are several sources of systematic errors in the standard way of
inferring this from the Galactic neutral hydrogen column density,
$N_{\rm H}$. Firstly, we invariably make the assumption, except when
forced not to, that the neutral hydrogen is the only tracer of metals
in the Galaxy, although molecular hydrogen can be important at some
locations on the sky; secondly, we assume a fixed elemental abundance
in the absorbing material; and thirdly, we have to rely on radio-based
measurements of the column density which may not be accurate. The
first of these is not known to be a problem in the case of any of our
targets, and there is little we can do about the second. However, to
assess the degree to which the third is a problem, we used two
different widely used databases to estimate interpolated values for
the Galactic $N_{\rm H}$. These were the compilation of Dickey \&
Lockman (1990), as provided by the standard {\it Chandra} on-line
proposal planning tool COLDEN\footnote{See
  http://cxc.harvard.edu/toolkit/colden.jsp .}, and the more recent,
higher-resolution Leiden/Argentine/Bonn Galactic HI survey (Kalberla
\etal\ 2005) for which an on-line search tool is also
available\footnote{See http://www.astro.uni-bonn.de/\textasciitilde
  webaiub/english/tools\_labsurvey.php}. Both values are tabulated in
Table \ref{sources} and it will be seen that there are differences at
the level of 10--20 per cent for each source, which allows us to get
an idea of the potential systematic effects on the results. We comment
below on the effects of using the two different estimates of Galactic
$N_{\rm H}$ in our spectral fitting.

\section{Methods}

In systems in which there is a clear excess of X-ray emission
associated with the lobe, estimating the inverse-Compton flux and, if
possible, measuring its spectrum is relatively easy, using local
(off-lobe) background regions to account for contamination by thermal
emission from the regions in front of and behind the lobe (as done by,
e.g., Hardcastle \& Croston 2005). By doing this we implicitly assume
that any difference between the surface brightness of thermal emission
on and off-lobe may be neglected; this is a safe assumption if the
lobe-related emission is clearly dominant.

The situation is very different in our chosen targets. Here it is no
longer safe to assume that the variation in the surface brightness of
the thermal emission can be neglected; there may well be visible
`cavities' in the thermal emission associated with the lobes, as
already seen in Cygnus A (Wilson \etal\ 2006) and Hydra A (Wise
\etal\ 2007), in which case na\"\i ve local
background subtraction would give negative fluxes for any
inverse-Compton emission. More problematically, there may be
variations in the temperature or abundance of the thermal emission on
scales comparable to those of the lobe. Failure to account for these
can in principle give spectral differences between the lobe and
off-lobe regions that can mimic the power-law spectrum of
inverse-Compton emission. On the other hand, accounting for them by
allowing the fitted temperature or abundance of the thermal emission
coincident with the lobe to vary freely will almost certainly reduce
or remove any spectral evidence for additional inverse-Compton
emission. In assessing the implications of any spectral detection or
non-detection of inverse-Compton emission it is necessary to
understand (1) whether the method being employed is actually capable
of detecting inverse-Compton if it is present, (2) whether it is
biased and what the uncertainties are on any detection, and (3) what
upper limits on inverse-Compton emission can be assigned in the case
of a non-detection.

To investigate these problems we carried out extensive {\sc xspec}
simulations for each source. We first defined spectral extraction
regions appropriate to the lobes of the radio sources of interest.
These consisted of the lobe region itself; two adjacent, identically
sized regions (`off-lobe' regions) intended to allow us to estimate
the local properties of thermal emission; and an `off-source'
background region for which the surface brightness of thermal emission
was much less than that in the on-source regions (though for all of
these sources it is hard to select a local background region which is
free of any thermal emission) which was used as background for all
three on-source regions. We extracted spectra from these regions using
the appropriate {\sc ciao} and {\sc sas} tools, and by fitting {\sc apec}
models to measure the temperature and abundances in the three
on-source regions, assuming that they can be described by a
  single temperature, we were able to characterise the properties of the
thermal emission near to the lobes. (Details of the fitted results for
the target sources are discussed in the next section of the paper.)

Using these fits we were then able to simulate, within {\sc xspec},
realistic observations of sources with comparable properties, using
the response files determined from the real datasets. Our simulations
included a spectrally realistic background (modelled using fits to the
real background data) and the simulated data were
background-subtracted and grouped in exactly the same way as for the
real data. The normalizations of the thermal ({\sc apec}) models used
for the three on-source regions, and their metal abundance (assumed
constant, since spatial variations in abundances for our targets,
though non-zero where they have been investigated, are not large on
the relevant scales: see Simeonescu \etal\ (2009) and Smith
\etal\ (2002) for the cases of Hydra A and Cygnus A respectively),
were based on the real data, so that the statistics were guaranteed to
be realistic. However, the temperatures of the thermal models were not
fixed for each simulation, but drawn from a Gaussian distribution with
a mean set equal to the temperature determined from a joint fit to the
two off-lobe regions and with a dispersion estimated from the standard
deviation of the two individually fitted temperatures; this was
intended to take account both of the uncertainty in the fitted
temperatures and of the possibility of variation in the temperature as
a function of the position in the cluster. In addition, a power-law
component with a fixed photon index (for these simulations we
initially used 1.5; see below for more discussion of this point) and a
variable normalization was added to the simulated data for the lobe
region. For a non-zero normalization of this component, we could
investigate whether a given method for estimating the level of
inverse-Compton could recover an unbiased estimate of the
normalization and estimate confidence limits on fits. If the
normalization was set to zero, the simulations allowed us to estimate
upper limits for non-detections.

We emphasise that this method is mainly designed to assess the effect
on the power-law normalization of point-to-point variations in the
temperature and normalization of a {\it single-temperature} model.
Taking account of possible multiple temperatures within a region would
be possible, but cumbersome, and clearly impossible to do in full
generality, since the regions of interest could contain an arbitrary
number of different components. However, the uncertainties on the
power-law normalization should be reasonably accurate to the extent
that the orginal models used are reasonable representations of the
off-lobe regions, i.e. that the fit of a single-temperature model is
reasonably good. We do not require a thermal model to represent the
cluster physics exactly so long as it characterizes the overall
spectral shape and its variations adequately.

Within this framework we investigated four possible methods for
estimating the level of inverse-Compton emission from the lobe regions
from spectral fitting. In all four, the basic approach was to fit
models consisting of thermal emission only to the two off-lobe regions
and a model consisting of thermal emission plus a power-law component
with fixed $\Gamma = 1.5$ to the lobe region, with the thermal
normalizations of all regions free to vary in order to take proper
account of the known or probable cavities in the X-ray emission
associated with the radio lobes. The models differed in the way in which
they related the properties of the thermal emission from the off-lobe
and lobe regions, as summarized below:

\begin{enumerate}
\item Abundances for lobe and off-lobe regions are tied together, but
  the on-lobe temperature is free to vary.
\item Both abundances and temperatures for the lobe and off-lobe
  regions are tied together (i.e. we fit a single temperature and
  abundance for the whole region).
\item The two off-lobe regions are allowed to have different
  temperatures, and the temperature fitted to the lobe region is
  constrained to be the mean of the temperatures fitted to the two
  off-lobe regions (the abundance is jointly derived from the two
  off-lobe regions; as in practice little abundance variation is seen,
  a model in which the mean abundance was used would be similar).
\item A single temperature and abundance are determined from a joint fit to the
  off-lobe regions only, and then frozen in a separate fit to the lobe
  region (i.e. only the normalization of the thermal and non-thermal components are free
  to vary in the fit to the lobe region).
\end{enumerate}

Clearly these four models represent different tradeoffs between, on
the one hand, the necessity to constrain some aspect of the lobe model
from the off-lobe data and, on the other, the recognition that
properties of the thermal emission may be different for lobe and
off-lobe regions. The simulations showed, however, that only the most
restrictive model (model iv) allowed an unbiased recovery of low-level
simulated power-law emission. In models (i)-(iii), the normalization
of the recovered power-law component was always systematically low,
presumably because either the abundance or the temperature of the
dominant thermal model tended to vary so as to account for some of the
simulated power law. Model (iv) is unbiased, even in the presence of
quite large scatter in the simulated temperature for the three
regions\footnote{One can imagine that model (iv) would be biased in a
  situation in which the thermal component of the lobe spectrum had a
  temperature, abundance or normalization that was {\it
    systematically} biased with respect to the off-lobe spectra, but it is
  essentially impossible to deal with this situation with the existing
  data, so we must simply make the assumption that this is not the
  case in order to proceed.}, although of course larger uncertainties
on the true temperature in the lobe region result in larger scatter on
the inferred power-law normalization. In what follows, therefore, we
use model (iv) to estimate the normalization of the power-law
component in fits to the real data, and we interpret the results using
the confidence limits on power-law normalization derived from
simulations.

Finally, we investigated the results of using a steeper photon index
for the simulated power laws. As we will discuss below, it is not
clear what the appropriate photon index for inverse-Compton models is.
We found that steeper photon indices produce systematically higher
upper limits on the 1-keV power-law normalization. This is presumably
because the difference between thermal and non-thermal models is most
obvious at the highest energies: softer non-thermal models are more
easily concealed by thermal emission. Although we use the results of
the simulations with $\Gamma = 1.5$ in what follows, we comment where
appropriate on the effects of a steeper photon index.

\section{Results}

For each of the three sources we extracted spectra and carried out
spectral fits for the on- and off-source regions shown in
Fig.\ \ref{picture} and described in the following subsections. The
results of the fits (for both values of $N_{\rm H}$) are tabulated in
Table \ref{results}. Errors quoted are 1$\sigma$ statistical errors
except where otherwise stated. Constraints on the particle content as
a function of assumptions about the electron energy spectrum are
summarized in Table \ref{kappa-summary}.

\begin{table*}
\caption{Results of spectral fitting. This table gives the
  best-fitting temperatures and abundances for the off-lobe regions
  together with the $\chi^2$ and number of degrees of freedom for the
  joint fit to these regions, the normalizations for the additional
  power-law components fitted to the lobe together with the fitting
  statistic for the lobe region, and the 99 per cent confidence upper
  limits on normalization derived from simulations, for each source
  region and for both the NRAO and LAB values of the Galactic $N_{\rm
    H}$. Note that as the results for both lobes of Her A are
  identical the table has only one entry for these.}
\label{results}
\begin{tabular}{llrrrrrrr}
\hline
Source&Region&$N_{\rm
  H}$&\multicolumn{3}{l}{Off-lobe}&\multicolumn{3}{l}{Lobe}\\
&&used&kT (keV)&Abundance&$\chi^2$/dof&1-keV flux
(nJy)&$\chi^2$/dof&1-keV flux limit (nJy)\\
\hline
Hydra A&Bubble&NRAO&$3.56 \pm 0.04$&$0.28 \pm 0.02$&2282/1944&0.6&1107/964&$<39$\\
&&LAB&$3.67 \pm 0.04$&$0.31 \pm 0.02$&2243/1944&0&1070/964&$<42$\\
Her A&Lobe&NRAO&$4.9 \pm 0.1$&$0.48 \pm 0.06$&528/580&0&321/327&$<38$\\
&&LAB&$4.9 \pm 0.1$&$0.48 \pm 0.06$&535/580&0&328/327&$<38$\\
Cyg A&W lobe&NRAO&$5.13 \pm 0.05$&$0.91 \pm 0.03$&3340/3128&82&1624/1675&$<66$\\
&&LAB&$6.26 \pm 0.05$&$1.02 \pm 0.03$&3892/3128&72&1758/1675&$<66$\\
&E lobe&NRAO&$4.33 \pm 0.03$&$0.96 \pm 0.03$&3926/3619&185&2443/2218&$<100$\\
&&LAB&$5.07 \pm 0.03$&$1.07 \pm 0.03$&4733/3619&178&3005/2218&$<100$\\
\hline
\end{tabular}
\end{table*}

\subsection{Hydra A}

For Hydra A we extracted spectra for and around the large-scale bubble
seen in low-frequency radio observations (Fig.\ \ref{picture}), 
  as this seemed to us to be the region of the source that was most
  likely to be bright in inverse-Compton emission compared to its
  surroundings.

The best-fitting temperatures for the off-lobe, cluster regions around
3.6 keV and abundances around 0.30 solar (Table \ref{results}) are in
reasonably good agreement with the properties of the ICM at this
distance from the nucleus obtained by Simionescu
\etal\ (2009) using a more complex thermal model. Fixing the
temperature and abundance for the lobe region and allowing the
normalization of the power law to vary, we obtain a power-law
normalization consistent with
zero within the errors derived from the fit for both values of $N_{\rm
  H}$. Using our {\sc xspec}
simulations, we derive a 99 per cent confidence upper limit on the
power-law normalization corresponding to 39
nJy at 1 keV for the NRAO $N_{\rm H}$ values: the values from the
alternative $N_{\rm H}$ value are similar.

We can now ask what this implies for the electron population. Clearly
the number density of the population of electrons around $\gamma \sim
1000$ cannot be so high as to cause the inverse-Compton emission to
exceed the limit we have derived. For a given assumption about the
low-energy ($\gamma < 1000$) electron energy spectrum, which is not
constrained by any observation, we can then put an upper limit on the
amount of the internal pressure of the lobe provided by electrons. The
low-energy electron energy spectrum is parameterized by the electron
energy index $p$, such that $N(E) = N_0 E^{-p}$, and the low-energy
Lorentz factor cutoff, $\gamma_{\rm min}$, and the ratio of the energy
density in non-radiating particles and in electrons is given by the
factor conventionally denoted $\kappa$. Making the assumptions that
the lobe is a sphere uniformly filled with electrons, non-radiating
particles and magnetic field such that (1) the observed synchrotron
spectrum is reproduced, (2) the external thermal pressure is equal to
the internal pressure due to the particles and field (this is a
conservative assumption; the internal pressure cannot be less than the
external pressure), and (3) the lobe
is not magnetically dominated, we can then ask what values of $p$,
$\gamma_{\rm min}$ and $\kappa$ are consistent with the
inverse-Compton upper limit. To do this we used the inverse-Compton
code of Hardcastle, Birkinshaw \& Worrall (1998a), which numerically
integrates the results of Rybicki \& Lightman (1979) for an arbitrary
electron energy spectrum given a spherical source geometry, taking into account
scattering of both the microwave background and synchrotron photons.
We took the external pressure to be $3.2 \times 10^{-12}$ Pa, based on
our measured temperature and an external density derived from the
deprojection of David \etal\ (2001) for a radius of 3.5 arcmin.

We initially limited the parameter space to be explored by making the
conservative assumption that $\gamma_{\rm min} = 1$. Then for $p=2$ (a
conventional choice based on the assumption of first-order Fermi
particle acceleration) the upper limit on inverse-Compton emission
becomes a lower limit on $\kappa$: we find that $\kappa > 16$ for
these choices. In other words, for these assumptions, the electrons
contribute $<1/17$ of the energy density in the lobe, with the balance
being made up almost entirely of non-radiating particles. (Assumption
(3) above means that the magnetic field energy density is always
negligible in comparison to the particle energy density, though for
very large values of $\kappa$ it starts to exceed the electron energy
density.) Increasing $\gamma_{\rm min}$ decreases the total electron
energy density further: for $\gamma_{\rm min} = 10$ the electrons
contribute $<1/23$ of the energy density in the lobe.

Higher values of $p$ rapidly increase the total electron energy
density for a given constraint at $\gamma \sim 1000$. So with $p =
2.5$, $\gamma_{\rm min} = 1$ we find that the upper limit on the
electron energy density is $<1/3$ of the total, falling to $<1/8$ for
$\gamma_{\rm min} = 10$. For $p=3.0$, the inverse-Compton emission
provides no constraint for $\gamma_{\rm min} = 1$ but restricts the
electron energy density contribution to $\la 1/2$ the total for
$\gamma_{\rm min} = 10$. These limits are in fact less strong than
they appear, since a choice of $p>2.0$ means that the inverse-Compton
photon index should be $>1.5$, which, as discussed above, increases
the limits on the power-law normalization and so the 1-keV flux, not
taken into account in the above calculations. However, the flat
spectral index ($\alpha = 0.73$) measured at high frequencies at the
base of the lobes in Hydra A (Taylor \etal\ 1990) is inconsistent with
$p > 2.5$, while it has been argued (Young \etal\ 2005) that $p
\approx 2.0$ is the correct value to adopt in FRI sources. We consider
that the results for $p = 2.0$ are the ones most likely to reflect
reality.

\subsection{Her A}

For Her A we investigated both the E and W lobes. Because of the
symmetry of the source we were able to use identical off-lobe regions
for both lobes. There is a clear detection of the E jet in this {\it
  Chandra} observation, corresponding to the radio structures denoted
E5, E6 and E11 by Gizani \& Leahy (2003), which we mask out in our
observation of the E lobe.

The best-fitting temperature for the off-lobe regions was $4.9 \pm
0.1$ keV, with abundance $0.48 \pm 0.06$ solar, which is consistent
with the earlier {\it Chandra} measurements of Nulsen \etal\ (2005),
bearing in mind that our regions necessarily span the strong surface
brightness drop described in that paper. For both lobes and for both
$N_{\rm H}$ values, a fit with fixed temperature and abundance gave a
best-fitting power-law normalization of zero. There is thus no
spectroscopic evidence for inverse-Compton emission in this source
despite the clear elongation of the X-ray emission in the direction of
the lobes. The 99 per cent upper limit on the inverse-Compton flux for
both lobes that we derive from our {\sc xspec} simulations corresponds
to a flux density at 1 keV of 38 nJy, independent of the choice of
$N_{\rm H}$ value. This comparatively large value
is a result of the large difference (nearly 0.6 keV) between the
best-fitting temperatures for the two off-lobe regions when fitted
separately.

Using the deprojection of Nulsen \etal\ (2005) we estimate the
pressure at the radius of the lobes, 1 arcmin, to be $7 \times
10^{-12}$ Pa. (This is the pressure just inside the surface
brightness/density jump.) Applying the same method as used for Hydra
A, we find that the electrons can contribute at most $1/4$ of the
total particle energy density for $p=2$, $\gamma_{\rm min} = 1$, and
around $1/6$ for $\gamma_{\rm min} = 10$. For $p \ge 2.5$ with
$\gamma_{\rm min} = 1$, the inverse-Compton limit imposes no
constraint on the fraction of internal energy contributed by the
electrons (i.e. they can provide all the required internal pressure
without violating the limit). As the flattest-spectrum structure in
the jet has $\alpha \approx 0.6$ (Leahy \& Gizani 2003) there is some
evidence that $p$ is likely to be close to 2.0, assuming that most
particle acceleration takes place in the jet and that the injection
spectral index is flat. The X-ray detection of the jet supports this
model in the sense that it suggests that particle acceleration does
take place in the jet, assuming a synchrotron origin for the X-rays.

We note that the strong depolarization asymmetry (Gizani \& Leahy
2003) implies that the source is some way from the plane of the sky
(Gizani \& Leahy 2004 estimate an inclination angle of $40^\circ$ to
the line of sight). Including projection in our analysis would
decrease the external pressure but increase the expected
inverse-Compton emissivity (since the lobes would be larger) so the
limits on the contribution of electrons to the total energy density
would remain roughly similar. Detailed modelling of the cluster
density profile and the constraints on projection angle imposed by the
observed depolarization should be possible with existing data, but is
well beyond the scope of the present paper.

\subsection{Cygnus A}
\label{cyga}

For Cygnus A we extracted spectra for both radio lobes, masking out the
well-known X-ray detected hotspots (e.g. Harris \etal\ 1994).
Overlapping but independent off-lobe regions were used for the two
lobes. There is evidence for temperature variation within the lobe and
off-lobe regions in discrete features (Wilson, Smith \& Young 2006;
Belsole \& Fabian 2007)
but the contribution to the spectrum of these features, given their
low surface brightness contrast, is expected to be minimal and we do
not attempt to exclude them.

For the W lobe the derived temperatures and abundances for the NRAO
$N_{\rm H}$ are very similar to those expected from the radial
profiling of earlier work (e.g. Smith \etal\ 2002) (our regions extend
between about 0.2 and 1 arcmin from the nucleus). The best-fitting
temperatures are significantly higher using the lower LAB column
density, but the $\chi^2$ values are also significantly worse,
consistent with the fact that Smith \etal\ fitted a $N_{\rm H} \sim
3.1 \times 10^{20}$ cm$^{-2}$. Here the addition of a power-law model
gives a well-constrained non-zero normalization corresponding to 82
nJy (NRAO) or 72 nJy (LAB), well above the detection threshold from
simulations. Deriving 90 per cent confidence error bars from
simulation, we can claim a detection with a flux of $82_{-60}^{+50}$
nJy at 1 keV (here and hereafter we only use the values corresponding
to the NRAO $N_{\rm H}$ value, noting that the results are similar for
the LAB $N_{\rm H}$, so that we are not dominated by systematic errors).

For the E lobe we again find an apparent detection of power-law
emission, this time with a flux density of $185_{-120}^{+60}$ nJy
again with 90 per cent confidence error bars derived from simulation).
For this lobe, the power-law component represents a little less than
half of the total counts seen in the spectrum. In both cases, of
course, we are fitting a power-law model with a fixed photon index of
1.5, and consequently we cannot rule out the possibility that the
second spectral component is a hot thermal component (with $kT \ga 10$
keV) but we neglect this possibility in what follows since we have an
a priori reason to suppose that we might see a hard power-law
component in Cyg A. The effect of reducing the assumed $N_{\rm H}$ is
seen (Table \ref{results}) to be, as expected, to reduce the inferred
flux density in the power law, but in neither case does it render the
detection insignificant.

Given the very large error bars (which arise because of the large
amount of temperature structure in Cyg A: essentially the simulations
are taking into account the fact that we do not know the true
temperature associated with the thermal emission from the lobes) the
fluxes of the two lobes are consistent with each other. The fact that
the E lobe appears brighter may partially be explained by the fact
that this lobe contains a relatively bright linear feature which (in
spectral fitting with a power-law model using local background
subtraction) could account for around 28 nJy of the power-law flux.
However, as the flux of this feature is negligible within the
uncertainties on our measurement, we do not correct for it in our
analysis.

We briefly investigated whether the normalization we had estimated for
the power-law component could be affected by the presence of
temperature structure in the cluster. We considered the spectrum of
the western off-lobe region since the western lobe detection is
considerably the weaker of the two. Initially we noted that when we
fitted the temperature of these two regions with a free Galactic
$N_{\rm H}$ the value we obtained was $3.1 \times 10^{21}$ cm$^{-2}$;
this is consistent with the results of Smith \etal\ (2002) and very
similar to the NRAO value. This is encouraging since spectra where
multi-temperature effects are significant are often best fitted with
$N_{\rm H}$ values significantly below the Galactic value. Fixing
$N_{\rm H}$ to the NRAO value again, we next fitted the spectra of the
western off-lobe regions with a model consisting of the sum of two
APEC models, with tied but free abundances and completely free
temperatures and normalizations. This gave a moderate improvement in
the fit to the spectrum ($\chi^2 = 3212$ for 3125 degrees of freedom:
compare results for single-temperature fits in Table \ref{results})
with $kT_1 = 2.78_{-0.17}^{+0.46}$ keV, $kT_2 = 7.44_{-0.38}^{+0.80}$
keV, abundance $0.98 \pm 0.04$ solar. However, when we fitted this
model, with fixed temperatures and abundances but free normalizations
for the two temperatures, together with an additional power-law
component, to the spectrum of the W lobe, the fitting statistic was
basically identical to that for the single-temperature model ($\chi^2
= 1628$ for 1674 d.o.f.) and the power-law normalization was
essentially unchanged (71 nJy at 1 keV). Similar results were found
for the E lobe. We conclude that temperature structure of this type
has little effect on our results, though of course a strongly
spatially varying soft component with low abundance could mimic our
power-law detection.

We therefore considered the implications of these apparent detections
of inverse-Compton emission for the particle content of the source. We
take the external electron densities estimated for the Cygnus A lobes
by Dunn \& Fabian (2004) which, together with our temperatures, give
an external pressure of $6 \times 10^{-11}$ Pa at the mid-point of the
lobes, $\sim 40$ arcsec from the core [this agrees with earlier
  estimates, such as those by Hardcastle \& Worrall (2000) and Wilson
  \etal\ (2006), bearing in mind that there is nearly an order of
  magnitude variation in external density, and therefore pressure,
  along the lobes]. This allows us to model the lobes in the same way
as for the previous two sources. However, there are two additional
complications. One is that the photon energy density in the bright
lobes of Cyg A is dominated by synchrotron photons, not CMB photons;
therefore the synchrotron self-Compton process is dominant, and the
code of Hardcastle \etal\ (1998a), which assumes spherical symmetry,
cannot give exact answers. We have verified using a more sophisticated
inverse-Compton code (Hardcastle \etal\ 2002), which by modelling the
source using an arbitrarily fine grid can account for the
self-illumination of any source geometry at the cost of greatly
increased computing time, that the correction for a uniform
ellipsoid is only of the order of 10 per cent, which is negligible
given the errors in the measured flux density. Secondly, here we have
a detection, rather than an upper limit, to deal with; this means
that, assuming pressure balance, we can actually measure one of the
free parameters ($\kappa$, $p$, etc) rather than setting limits on it.

Both lobes are brighter than the prediction for IC at equipartition
for $\kappa = 1$, $\gamma_{\rm min} = 1$ and $p=2$, which is around 20
nJy. Thus here we already know, without considering external pressure
constraints, that we expect the electron energy density to dominate
over that of the field. However, the external pressure we use here
exceeds by almost an order of magnitude the minimum pressure in the
lobes, which is of the order of $4 \times 10^{-12}$ Pa for $p=2$
[similar results have been found by Hardcastle \& Worrall (2000) and
  by Dunn \& Fabian (2004)]. If we assume $p = 2$, $\gamma_{\rm min} =
1$ and $\kappa=0$ then the electron density required if the lobes are
to be in pressure balance gives inverse-Compton emission that exceeds
the observed value by a factor 5.5 (W lobe) to 2.4 (E lobe), and
cannot be consistent even given the large errors; thus we cannot
supply the missing pressure with electrons alone for this $p$ value.
To make the prediction consistent with observation we require $\kappa
\approx 4$ (W lobe) or 1 (E lobe), implying that non-radiating
particles are either in rough energy equipartition with the electrons
or dominant by a small factor. However, only modest increases in $p$
are needed to restore pressure balance and match the observed level of
inverse-Compton emission ($p=2.2$ -- $2.3$). In these models the
electrons still dominate the magnetic field energy density by a large
factor but no non-radiating particles are required. For $p \approx
2.1$ we could accommodate the inverse-Compton observations, obtain
pressure balance and have $\kappa = 1$. But unlike the situation in
some FRIIs in poorer environments (e.g. Hardcastle \etal\ 2002,
Croston \etal\ 2004), there does not appear to be a possible sitation
in which $\kappa = 0$, $p=2$ and both the inverse-Compton and pressure
balance constraints are satisfied with $B$-fields close to the
equipartition value if we assume that the sources is close to the
plane of the sky. Projection would have some effect, but, as discussed
above (and see Hardcastle \& Worrall 2000 for more) the effect is not
likely to be that large, especially for a source, like Cygnus A, where
the angle to the line of sight is likely to be $\ga 45^\circ$.

Finally, it is worth commenting on the assumption of pressure balance
in the case of a powerful FRII like Cygnus A. Our implicit assumption
throughout the discussion here and in other papers is that the lobes
are characterised by a single pressure, with a high internal sound
speed in the lobes acting to smooth out any pressure differences on a
timescale that is short compared to the dynamical time of the source.
At the same time, FRIIs in general, and Cygnus A in particular, have
external pressures that necessarily vary along the length of the
lobes; in Cygnus A's case, as noted above, this variation is an order
of magnitude in pressure terms. When discussing pressure balance of
the lobes we are in fact talking about pressure balance at (roughly)
the mid-point of the lobe; the outer edge of the lobe in this case
would have to be expanding and the inner edge contracting or,
equivalently, moving outwards under buoyancy [the `cocoon crushing'
  process discussed by Williams (1991) and Hardcastle \& Worrall
  (2000)] if the midpoint were in pressure balance. In fact, there is
some evidence for mildly supersonic expansion even at the midpoint of
the lobes of Cygnus A in the shape of the weak shocks argued for by
Wilson \etal (2006). If these are shocks, then the pressure in the
lobes should be equal to the pressure in the shocked gas, and using
the values for the mass density in this region quoted by Wilson
\etal\ in fact gives densities a factor 2 lower than the value from
Dunn \etal , although clearly any inference of density in these
regions is very geometry-dependent. The key point, though, is that
uncertainty is introduced into our constraints on particle content by
the uncertain dynamical state of the lobes of the source. If we
required the lobe to be expanding supersonically in all directions
(which is certainly {\it not} supported by the observations of Wilson
\etal\ (2006)) then we would need to use external pressures
corresponding to the central regions of the cluster, which are a
factor of a few higher, in our calculations above, leading to
correspondingly more extreme constraints on $\kappa$ and on the
fraction of the lobe energy density contributed by the electron
population.

\begin{table}
\caption{Summary of constraints described in the text on the fraction of the particle energy
  provided by electrons for each source as a function of $p$ and
  $\gamma_{\rm min}$. The numbers tabulated are equal to $1/(1+\kappa)$.}
\label{kappa-summary}
\begin{tabular}{lrrrr}
\hline
Source&$\gamma_{\rm min}$&\multicolumn{3}{c}{Constraints on electron
  energy content}\\
&&$p=2.0$&$p=2.5$&$p=3.0$\\
\hline
Hydra A&1&$<1/17$&$<1/3$&--\\
&10&$<1/23$&$<1/8$&$<1/2$\\
Her A&1&$<1/4$&--&--\\
&10&$<1/6$&$<1/2$&--\\
Cyg A&1&$1/5$ -- $1/2$&1&--\\
\hline
\end{tabular}
\end{table}

\section{Discussion and conclusions}

The results of the previous section may be summarized as follows:

\begin{itemize}
\item Constraints on the inverse-Compton emission from the lobes, on
  the assumption that the lobes are in pressure balance with the
  external medium, that they are uniformly filled with particles and
  field, and that they are not magnetically dominated, set limits on
  the fraction of the lobe energy density provided by electrons.
\item Because the lobe internal pressure is determined by the
  integrated electron energy spectrum, and not its normalization at
  $\gamma \approx 1000$, the results depend on the slope of the
  (assumed power-law) low-energy electron spectrum, $p$, and its
  low-energy cutoff, $\gamma_{\rm min}$. Increasing $\gamma_{\rm min}$
  reduces the lobe internal pressure for a given IC emissivity, while
  increasing $p$ increases it. Our fiducial assumptions are $p=2.0$,
  $\gamma_{\rm min}=1$ (though the latter neglects the possibility of
  Coulomb losses against any thermal particles present in the lobes, a
  process thought to operate effectively in supernova remnants; see,
  e.g., Vink 2008).
\item For a given choice of $p$ and $\gamma_{\rm min}$, and a value of
  $\kappa$, the ratio between the energy densities of radiating and
  non-radiating particles, there are only two possible solutions for
  the magnetic field strength, $B$, that allow us to achieve pressure
  balance given the observed synchrotron emissivity in the radio
  (which sets the electron energy spectrum normalization). If we
  assume that particles are dominant, so that the lower value of $B$
  should always be the one chosen, we can then constrain $\kappa$ by
  comparing the predicted and actual inverse-Compton emissivity.
  Increasing $\kappa$ decreases the inverse-Compton flux, so an upper
  limit on inverse-Compton flux corresponds to a lower limit on
  $\kappa$.
\item As shown in Table \ref{kappa-summary}, for $p=2.0$, $\gamma_{\rm
  min} = 1$ we find $\kappa > 16$ for the large-scale lobes of Hydra
  A, $\kappa > 3$ for the lobes of Her A, and $\kappa=1$--$4$ for
  Cygnus A (where a statistically significant detection of excess hard
  emission from the lobes allows a measurement rather than a limit to
  be derived, although the errors are large). All these values of
  $\kappa$ decrease if the assumed $p$ is increased, but within the
  limits set on $p$ by radio observations $\kappa = 0$ is not possible
  for either Hydra A or Her A. In other words, the electrons must
  contribute a small fraction of the total internal energy density in
  the FRI Hydra A and the intermediate source Her A, with the bulk of
  it, assuming particle dominance, being provided by a non-radiating
  particle population. For the FRII Cygnus A, the electron content of
  the lobes is close to the value required to provide pressure balance
  with the external medium at the mid-point of the lobes without a
  dominant contribution from non-radiating particles, but is
  insufficient to drive a strongly supersonic lobe expansion.
\end{itemize}

It has of course been known for a long time that solutions with
$\kappa=0$ and $B=B_{\rm eq}$ do not appear to be viable in the jets,
lobes and plumes of FRI radio galaxies (e.g.\ Hardcastle \etal\ 2007
and references therein). In some cases the lack of detected
inverse-Compton emission has been used to rule out a model in which
the lobes are electron-dominated (Hardcastle \etal\ 1998b, Croston
\etal\ 2003) while the use of the external pressure and the radio
fluxes to estimate a quantity equivalent to $\kappa$ is widespread
(e.g. Dunn \& Fabian 2004). All previous attempts to estimate
$\kappa$ in FRIs, though, rely on some assumption about the magnetic field
strength, normally one equivalent to equipartition between the field
and the particle population. The novelty of our work is that it
demonstrates that inverse-Compton constraints can be used to remove
the necessity for this assumption; our values of $\kappa$ are not just
a quantity that depends on the ratio of minimum internal to external
pressures, but real measurements that incorporate all the available
information. For models corresponding to the lower limits on $\kappa$,
we have $U_{\rm non-rad} > U_{\rm e} \gg U_B$ for Hydra A and Her A,
though we emphasise that we have ruled out a priori magnetically
dominated models, and the non-detections in inverse-Compton provide no
additional evidence for that assumption.

An interesting feature of the existing constraints on pressure balance
in lobes has been the apparent difference between FRI and FRII sources
found in earlier work. As discussed in Section \ref{intro}, when the
magnetic field strengths are constrained using inverse-Compton
detections in FRIIs, the internal pressures are typically comparable
to the external pressures with no additional non-radiating particles
and only modest departures from equipartition (that is, solutions with
$U_{\rm e} > U_B$, $U_{\rm non-rad} =0$ ($\kappa = 0$) are implied,
although typically a contribution from non-radiating particles of
order $U_{\rm e}$ ($\kappa \approx 1$) cannot be ruled out). In
typical FRIs, by contrast, it seems that solutions with $U_{\rm
  non-rad} \gg U_{\rm e}$ ($\kappa \gg 1$) are required by the data. Our
result for Cyg A confuses this simple picture. Although we appear to
detect inverse-Compton emission, as in other FRIIs, and the implied
electron energy densities are quite close to what would be required to
provide pressure balance, unlike the case for the FRIs, the data do
require some non-radiating particles and/or a departure from our
fiducial electron energy spectrum in order to allow pressure balance
with the external environment.
We emphasise that in
general, and also in the specific case of Cyg A, there are large
uncertainties on inferences of electron energy densities from
inverse-Compton observations, so that the differences between Cygnus A
and other FRIIs may not be as significant as they appear here.
However, an intriguing possibility is that the apparent difference
could be related to the very much richer environment of Cyg A when
compared to other FRIIs that have been studied with inverse-Compton
observations. In FRIs, there is some evidence that the required
non-radiating particle population may be related to entrainment (and
subsequent efficient heating) of external material, since jetted
sources like Hydra A tend to have larger pressure deficits than lobed
ones like Her A (Croston \etal\ 2008). Conceivably the entrainment
rates of the jets close to the centre of the rich cluster environment
of Cygnus A are high enough that the process operating in FRIs is also
relevant there, so that there is an environmental dependence for lobe
particle content both in FRIs and in FRIIs. The entrainment model for
FRIs will be discussed in detail in a forthcoming paper (Croston \&
Hardcastle, in prep.). For FRIIs, detailed testing of this idea must
await observations capable of detecting inverse-Compton emission in
other sources in rich environments.

This work has demonstrated both the potential and the severe
difficulties of inverse-Compton studies of lobe particle content in
the presence of a hot thermal environment. The very large
uncertainties on the inferred parameters even for such a bright
inverse-Compton source as Cygnus A, with such good X-ray data, make it
clear that our technique is not likely to be applicable to less
sensitive studies of more distant luminous FRIIs in clusters. The best
results were obtained for Hydra A, where the bubbles are large and the
surface brightness of thermal bremsstrahlung reasonably low; there are
almost certainly other well-studied large radio sources for which
inverse-Compton constraints would yield useful information. However,
our work clearly highlights the current problems in making a {\it
  spectral} separation between inverse-Compton and thermal emission.
To pin down the electron content in lobes in general, we need the
ability to image in somewhat harder X-rays, above the cutoff in
thermal bremsstrahlung, with good sensitivity and moderate angular
resolution. Hard X-ray imaging instruments such as the {\it Nuclear
  Spectroscopic Telescope Array} ({\it NuSTAR}), or, failing that, the
{\it International X-ray Observatory} ({\it IXO}) will eventually make
this possible.

\section*{Acknowledgements}

MJH thanks the Royal Society for a research fellowship. We thank Wendy
Lane and Paddy Leahy for providing us with published radio maps of
Hydra A and Her A respectively. We also thank an anonymous referee for
comments which allowed us to make significant improvements to the
paper. This research has made use of the
NASA/IPAC Extragalactic Database (NED) which is operated by the Jet
Propulsion Laboratory, California Institute of Technology, under
contract with the National Aeronautics and Space Administration. This
work is partly based on observations obtained with {\it XMM-Newton},
an ESA science mission with instruments and contributions directly
funded by ESA Member States and the USA (NASA). The National Radio
Astronomy Observatory is a facility of the National Science Foundation
operated under cooperative agreement by Associated Universities, Inc.

\end{document}